\documentstyle[twoside,epsf]{article}

\catcode`\@=11
\long\def\@makefntext#1{
\protect\noindent \hbox to 3.2pt {\hskip-.9pt  
$^{{\eightrm\@thefnmark}}$\hfil}#1\hfill}		

\def\thefootnote{\fnsymbol{footnote}}
\def\@makefnmark{\hbox to 0pt{$^{\@thefnmark}$\hss}}	
        
\def\ps@myheadings{\let\@mkboth\@gobbletwo
\def\@oddhead{\hbox{}
\rightmark\hfil\eightrm\thepage}   
\def\@oddfoot{}\def\@evenhead{\eightrm\thepage\hfil
\leftmark\hbox{}}\def\@evenfoot{}
\def\sectionmark##1{}\def\subsectionmark##1{}}

\oddsidemargin=\evensidemargin
\renewcommand{\thefootnote}{\fnsymbol{footnote}}

\newcounter{sectionc}\newcounter{subsectionc}\newcounter{subsubsectionc}
\renewcommand{\section}[1] {\vspace{12pt}\addtocounter{sectionc}{1} 
\setcounter{subsectionc}{0}\setcounter{subsubsectionc}{0}\noindent 
        {\tenbf\thesectionc. #1}\par\vspace{5pt}}
\renewcommand{\subsection}[1] {\vspace{12pt}\addtocounter{subsectionc}{1} 
        \setcounter{subsubsectionc}{0}\noindent 
        {\bf\thesectionc.\thesubsectionc. {\kern1pt \bfit #1}}\par\vspace{5pt}}
\renewcommand{\subsubsection}[1] {\vspace{12pt}\addtocounter{subsubsectionc}{1}
        \noindent{\tenrm\thesectionc.\thesubsectionc.\thesubsubsectionc.
        {\kern1pt \tenit #1}}\par\vspace{5pt}}
\newcommand{\nonumsection}[1] {\vspace{12pt}\noindent{\tenbf #1}
        \par\vspace{5pt}}

\newcounter{appendixc}
\newcounter{subappendixc}[appendixc]
\newcounter{subsubappendixc}[subappendixc]
\renewcommand{\thesubappendixc}{\Alph{appendixc}.\arabic{subappendixc}}
\renewcommand{\thesubsubappendixc}
	{\Alph{appendixc}.\arabic{subappendixc}.\arabic{subsubappendixc}}

\renewcommand{\appendix}[1] {\vspace{12pt}
        \refstepcounter{appendixc}
        \setcounter{figure}{0}
        \setcounter{table}{0}
        \setcounter{lemma}{0}
        \setcounter{theorem}{0}
        \setcounter{corollary}{0}
        \setcounter{definition}{0}
        \setcounter{equation}{0}
        \renewcommand{\thefigure}{\Alph{appendixc}.\arabic{figure}}
        \renewcommand{\thetable}{\Alph{appendixc}.\arabic{table}}
        \renewcommand{\theappendixc}{\Alph{appendixc}}
        \renewcommand{\thelemma}{\Alph{appendixc}.\arabic{lemma}}
        \renewcommand{\thetheorem}{\Alph{appendixc}.\arabic{theorem}}
        \renewcommand{\thedefinition}{\Alph{appendixc}.\arabic{definition}}
        \renewcommand{\thecorollary}{\Alph{appendixc}.\arabic{corollary}}
        \renewcommand{\theequation}{\Alph{appendixc}.\arabic{equation}}
        \noindent{\tenbf Appendix \theappendixc #1}\par\vspace{5pt}}
\newcommand{\subappendix}[1] {\vspace{12pt}
        \refstepcounter{subappendixc}
        \noindent{\bf Appendix \thesubappendixc. {\kern1pt \bfit #1}}
	\par\vspace{5pt}}
\newcommand{\subsubappendix}[1] {\vspace{12pt}
        \refstepcounter{subsubappendixc}
        \noindent{\rm Appendix \thesubsubappendixc. {\kern1pt \tenit #1}}
	\par\vspace{5pt}}

\newcommand{\textlineskip}{\baselineskip=13pt}
\newcommand{\smalllineskip}{\baselineskip=10pt}

\def\eightcirc{
\begin{picture}(0,0)
\put(4.4,1.8){\circle{6.5}}
\end{picture}}
\def\eightcopyright{\eightcirc\kern2.7pt\hbox{\eightrm c}} 

\newcommand{\copyrightheading}[1]
        {\vspace*{-2.5cm}\smalllineskip{\flushleft
        {\footnotesize International Journal of Modern Physics C, #1}\\
        {\footnotesize $\eightcopyright$\,\,\, World Scientific Publishing
         Company}\\
         }}

\newcommand{\publisher}[2]{{\begin{center}\footnotesize\smalllineskip 
        Received #1\\
        Revised #2
        \end{center}
        }}

\def\abstracts#1#2#3{{
        \centering{\begin{minipage}{4.5in}\baselineskip=10pt\footnotesize
        \parindent=0pt #1\par
        \parindent=15pt #2\par
        \parindent=15pt #3\par
        \end{minipage}}\par}} 

\newcommand{\bibit}{\nineit}
\newcommand{\bibbf}{\ninebf}
\renewenvironment{thebibliography}[1]
        {\frenchspacing
	 \ninerm\baselineskip=11pt
         \begin{list}{\arabic{enumi}.}
        {\usecounter{enumi}\setlength{\parsep}{0pt}     
         \setlength{\leftmargin 17pt}{\rightmargin 0pt}   
         \setlength{\itemsep}{0pt} \settowidth
	{\labelwidth}{#1.}\sloppy}}{\end{list}}

\newcounter{itemlistc}
\newcounter{romanlistc}
\newcounter{alphlistc}
\newcounter{arabiclistc}
\newenvironment{itemlist}
    	{\setcounter{itemlistc}{0}
	 \begin{list}{$\bullet$}
	{\usecounter{itemlistc}
	 \setlength{\parsep}{0pt}
	 \setlength{\itemsep}{0pt}}}{\end{list}}

\newcommand{\fcaption}[1]{
        \refstepcounter{figure}
	\setbox\@tempboxa = \hbox{\footnotesize Fig.~\thefigure. #1}
	\ifdim \wd\@tempboxa > 5in
           {\begin{center}
	\parbox{5in}{\footnotesize\smalllineskip Fig.~\thefigure. #1}
            \end{center}}
        \else
             {\begin{center}
	     {\footnotesize Fig.~\thefigure. #1}
              \end{center}}
        \fi}

\newcommand{\tcaption}[1]{
        \refstepcounter{table}
	\setbox\@tempboxa = \hbox{\footnotesize Table~\thetable. #1}
        \ifdim \wd\@tempboxa > 5in
           {\begin{center}
         \parbox{5in}{\footnotesize\smalllineskip Table~\thetable. #1}
            \end{center}}
        \else
             {\begin{center}
	     {\footnotesize Table~\thetable. #1}
              \end{center}}
        \fi}

\def\@citex[#1]#2{\if@filesw\immediate\write\@auxout
	{\string\citation{#2}}\fi
\def\@citea{}\@cite{\@for\@citeb:=#2\do
	{\@citea\def\@citea{,}\@ifundefined
	{b@\@citeb}{{\bf ?}\@warning
	{Citation `\@citeb' on page \thepage \space undefined}}
	{\csname b@\@citeb\endcsname}}}{#1}}

\newif\if@cghi
\def\cite{\@cghitrue\@ifnextchar [{\@tempswatrue
	\@citex}{\@tempswafalse\@citex[]}}
\def\citelow{\@cghifalse\@ifnextchar [{\@tempswatrue
	\@citex}{\@tempswafalse\@citex[]}}
\def\@cite#1#2{{$\null^{#1}$\if@tempswa\typeout
	{IJCGA warning: optional citation argument 
	ignored: `#2'} \fi}}

\def\pmb#1{\setbox0=\hbox{#1}
        \kern-.025em\copy0\kern-\wd0
        \kern.05em\copy0\kern-\wd0
        \kern-.025em\raise.0433em\box0}

\def\fnt#1#2{\footnotetext{\kern-.3em
        {$^{\mbox{\scriptsize #1}}$}{#2}}}

\def\fpage#1{\begingroup
\voffset=.3in
\thispagestyle{empty}\begin{table}[b]\centerline{\footnotesize #1}
        \end{table}\endgroup}

\def\runninghead#1#2{\pagestyle{myheadings}
\markboth{{\protect\footnotesize\it{\quad #1}}\hfill}
{\hfill{\protect\footnotesize\it{#2\quad}}}}
\font\tenbf=cmbx10
\font\tenit=cmti10 
\font\tenit=cmti10
\font\bfit=cmbxti10 at 10pt
\font\ninebf=cmbx9
\font\ninerm=cmr9
\font\nineit=cmti9

\font\eightrm=cmr8

\def\lsym{\raise-3pt\hbox{\vbox{\tabskip0pt\offinterlineskip
	\halign{\tabskip0pt plus 1em
	##\tabskip0pt\cr
	$\,\,<\,\,$\cr
	$\,\,\sim\,\,$\cr}}}}
\def\rsym{\raise-3pt\hbox{\vbox{\tabskip0pt\offinterlineskip
     \halign{\tabskip0pt plus 1em
      ##\tabskip0pt\cr
      $\,\,>\,\,$\cr
      $\,\,\sim\,\,$\cr}}}}
\def\qed{\hbox{${\vcenter{\vbox{			
	\hrule height 0.4pt\hbox{\vrule width 0.4pt height 6pt
	\kern5pt\vrule width 0.4pt}\hrule height 0.4pt}}}$}}
\def\theequation{\thesection.\arabic{equation}}		
\renewcommand{\thefootnote}{\fnsymbol{footnote}}	

\begin{document}

\runninghead{M. A. Novotny}
{Advanced Dynamic Algorithms $\cdots$}

\normalsize\textlineskip
\thispagestyle{empty}
\setcounter{page}{1}

\copyrightheading{Vol. 0, No. 0 (1999) 000--000}

\vspace*{0.88truein}

\fpage{1}
\centerline{\bf Advanced Dynamic Algorithms for the Decay of Metastable Phases}
\vspace*{0.035truein}
\centerline{\bf in Discrete Spin Models: Bridging 
Disparate Time Scales} 
\vspace*{0.37truein}
\centerline{\footnotesize M. A. Novotny} 
\vspace*{0.015truein}
\centerline{\footnotesize\it Supercomputer Computations Research Institute, 
Florida State University, Tallahassee, Florida 32306-4130, USA} 

\vspace*{0.225truein}
\publisher{24 August 1999}{24 August 1999}

\vspace*{0.21truein}
\abstracts{
An overview of advanced dynamical algorithms capable of
spanning the widely disparate time scales that govern the
decay of metastable phases in discrete spin models is presented.
The algorithms discussed include constrained transfer-matrix,
Monte Carlo with Absorbing Markov Chains (MCAMC), and
projective dynamics (PD) methods.  The strengths and weaknesses
of each of these algorithms are discussed, 
with particular emphasis on identifying the parameter regimes
(system size, temperature, and field) in which each algorithm works best.
}{}{}



\vspace*{1pt}\textlineskip	
\setcounter{section}{1}
\setcounter{equation}{0}
\section{Introduction}		
\vspace*{-0.5pt}
\noindent
One of the most challenging hurdles to be overcome before 
the goal of computational materials design can be realized is the problem 
of disparate time scales.  The time scales range from microscopic 
(typically an inverse phonon frequency, roughly $10^{-12}$~s) to the 
time over which devices made from the materials must operate 
(years to centuries).  The microscopic time scale is much 
smaller than the time between clock ticks on today's computers, and the 
times one desires to simulate are much too long to 
be achieved with traditional algorithms.  
Consequently, overcoming the hurdle of disparate 
timescales 
will require advanced dynamic algorithms which perform 
faster-than-real-time simulations.  Here we briefly review three algorithms 
that allow such simulations for metastable decay in 
discrete spin models.  Note that since we are interested in the dynamics 
of the problem, we cannot use advanced algorithms that change the 
dynamics (such as cluster algorithms, multicanonical algorithms, 
hybrid Monte Carlo algorithms, etc).  Each advanced dynamic algorithm 
has its strengths and weaknesses, so which one should 
be used will depend not only on the individual problem to be simulated, 
but also on the parameter regime in which the problem is to be studied.  

We concentrate on applying these advanced dynamical algorithms to 
the case of metastable decay of the square-lattice Ising model with 
periodic boundary conditions.  This simple model has been used to study 
the metastable state of nanoscale magnets\cite{GenMag} 
and the thermal decay of 
information stored on magnetic recording media\cite{EEMag}.  
The Hamiltonian is 
${\cal H}$$=$$-$$J$$\sum_{\langle i,j\rangle} 
\sigma_i \sigma_j$$-$$H \sum_i \sigma_i$, 
with the Ising spins $\sigma$$=$$\pm$$1$.  
The first sum is an interaction between nearest neighbors on a square 
$L$$\times$$L$ lattice, 
and the second sum is the interaction with the external field $H$.  
The temperature is taken to be below the critical temperature, 
$T_{\rm c}$$\approx$$2.26$$\cdots$$J$.  
We start with all spins 
up (equal to $+$$1$) and $H$ negative.  
In addition to the Hamiltonian, the dynamic must be specified:  
here the dynamic is randomly choosing one of the $L^2$ spins 
and deciding whether or not to flip it using a Metropolis 
flip probability\cite{MCbook}.  
One quantity we are interested in is the average lifetime, 
$\langle\tau\rangle$, 
before the system reaches some cut-off magnetization (taken to be zero) 
as it decays to the equilibrium state.  
Even in this simple model, which decays by homogeneous nucleation and 
growth, there are different regimes where $\langle\tau\rangle$ has 
different functional forms\cite{RTMS94}.  There are 
four relevant length scales in the problem (far below $T_{\rm c}$).  
Besides $L$ and the lattice spacing, $a$, these are the critical 
droplet radius $R_{\rm c}$ 
and the typical distance $R_o$ that a supercritical droplet (one with 
$R$$>$$R_{\rm c}$) 
grows before it encounters another supercritical droplet.  
The critical droplet radius for circular droplets is 
$R_{\rm c}$$=$$\sigma_\infty(T) / 2 |H| m_{\rm sp}$
where $\sigma_\infty(T)$ is the surface tension along a primitive lattice 
vector when $L$$\rightarrow$$\infty$, and 
$m_{\rm sp}$ is the spontaneous magnetization.  
For the square lattice both 
$\sigma_\infty(T)$ and $m_{\rm sp}$ are known exactly.  
The `cross-over dynamical phase diagram' 
for metastable decay is shown in Fig.~1.  

\textheight=7.7truein		
\setcounter{footnote}{0}
\renewcommand{\thefootnote}{\alph{footnote}}

\begin{figure}
\vspace*{13pt}
\vspace*{0.4in}
\centerline{
\epsfxsize=3.8in
\epsfysize=3.8in
\epsfbox{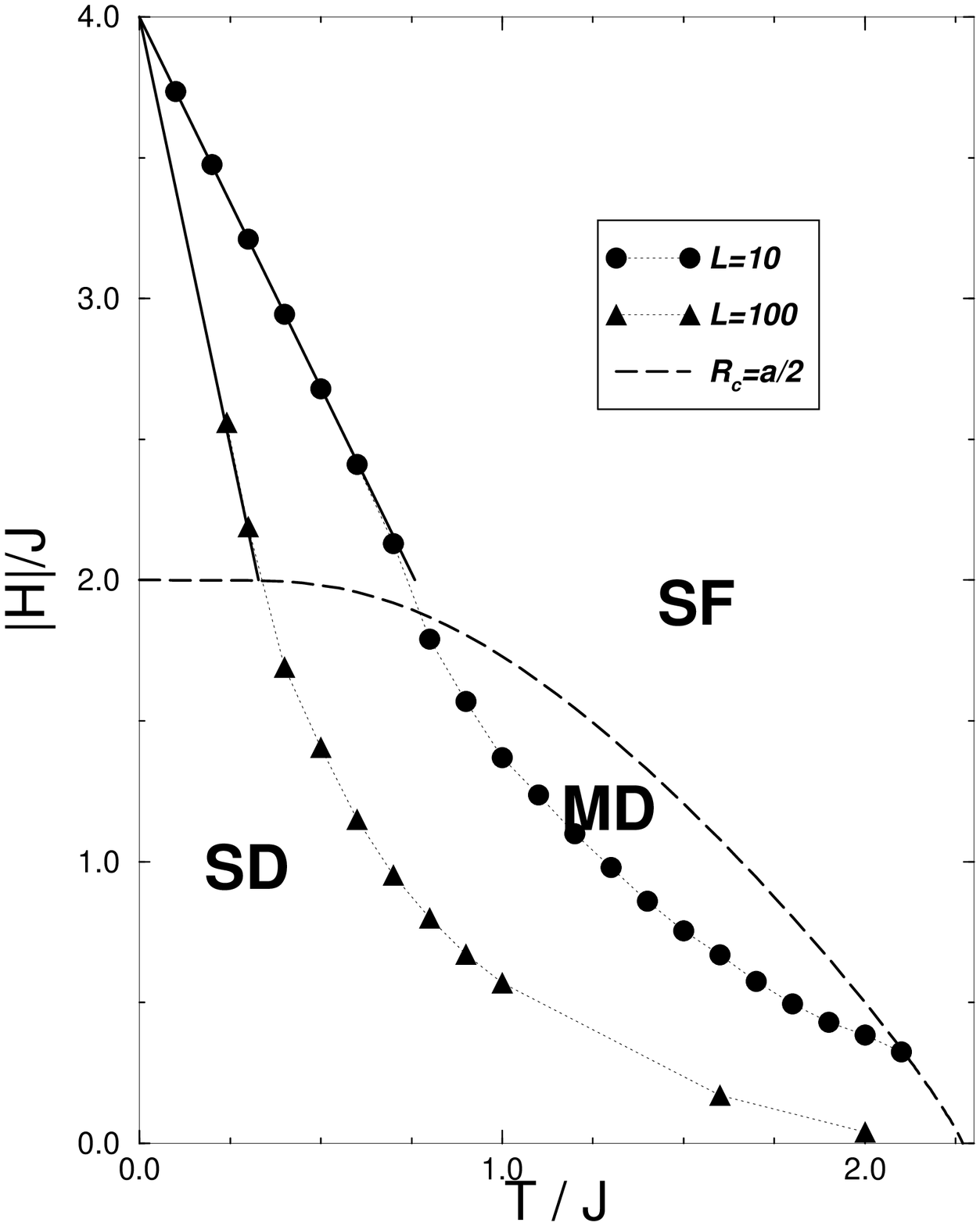}
}
\vspace*{10pt}
\fcaption{
The `cross-over dynamical phase diagram' for metastable decay of
the square-lattice Ising model.  
The dashed line corresponds to the
`mean-field spinodal'.  
The `dynamic spinodal'
is shown for $L$$=$$10$ and $L$$=$$100$.
This is a heavy solid line for the analytical expression for 
$|H|$$>$$2$$J$, and dotted lines which connect the data points.  
Shown are the Strong Field (SF),
Multi-Droplet (MD), and Single Droplet (SD) decay regimes.}
\end{figure}
\noindent 
For fields that are too strong the droplet picture of 
decay is not valid,  
this is the Strong Field (SF) regime\cite{RTMS94}.  
For weaker fields 
the nucleation of multiple droplets (MD regime)\cite{RTMS94} 
leads to decay of the metastable state via an Avrami 
mechanism\cite{Ramos99} 
since 
$R_{\rm c}$$\ll$$R_o$$\ll$$L$. 
A conservative estimate 
for the `mean-field spinodal' that separates the SF and MD regimes 
is given by $R_{\rm c}$$=$$a$$/$$2$, independent of $L$.  
For weak fields and low temperatures a nucleation of a 
single droplet (SD regime) leads to decay of the metastable state, 
since $R_{\rm c}$$\ll$$L$$\ll$$R_o$.  
Our estimate\cite{RTMS94} of the `dynamic spinodal' (where 
$R_o$$\sim$$L$) that gives the limit of the 
SD regime is where the standard deviation of the lifetime is 
equal to $\langle\tau\rangle$$/$$2$.  
This cross-over depends on system size, and Fig.~1 shows simulation 
values using $10^3$ escapes for $L$$=$$10$ and $L$$=$$100$.  
At $|H|$$>$$2$$J$ and 
very low temperatures a single overturned spin is the 
nucleating droplet\cite{LowT1} and the 
`dynamic spinodal' for our dynamic is given by\cite{LeeNR95}
$(4$$-$$H_{\rm DSp}$$)$$/$$T$$=$${3\over 2}$$\ln(L)$$-$$0.82$.  
Only the last term is an adjustable parameter, which 
has been chosen to 
give agreement for $L$ between $8$ and $240$.  
For other dynamics, such as sequential updates, the functional form for 
$H_{\rm DSp}$ is different\cite{LeeNR95}.  
The `coexistence regime'\cite{RTMS94}, 
with $L$$<$$R_{\rm c}$, is not shown in Fig.~1.  

\begin{figure}
\vspace*{13pt}
\vspace*{0.8in}
\centerline{
\epsfxsize=3.3in
\epsfysize=3.3in
\epsfbox{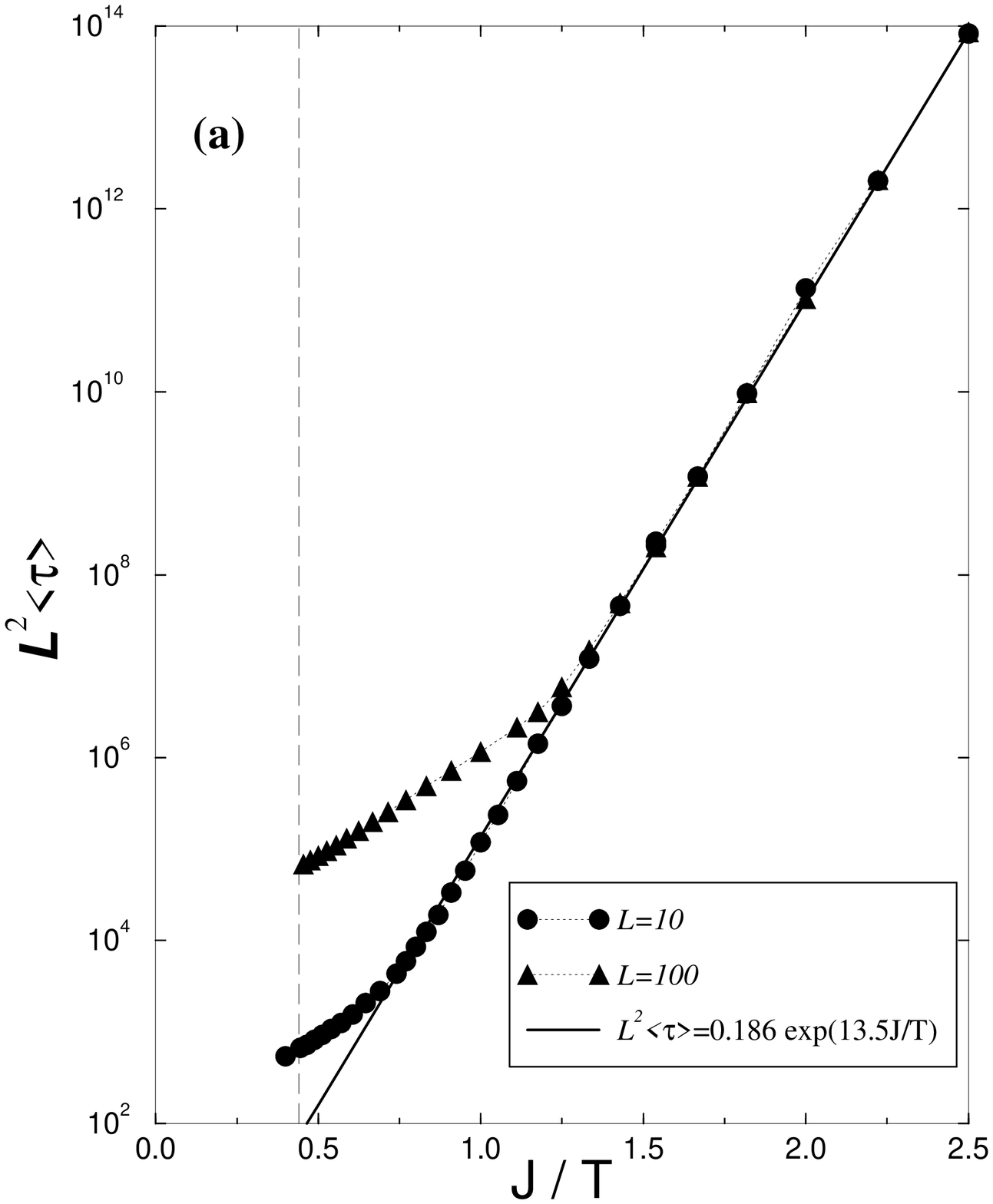}
\hspace*{-0.2true cm}
\epsfxsize=3.3in
\epsfysize=3.3in
\epsfbox{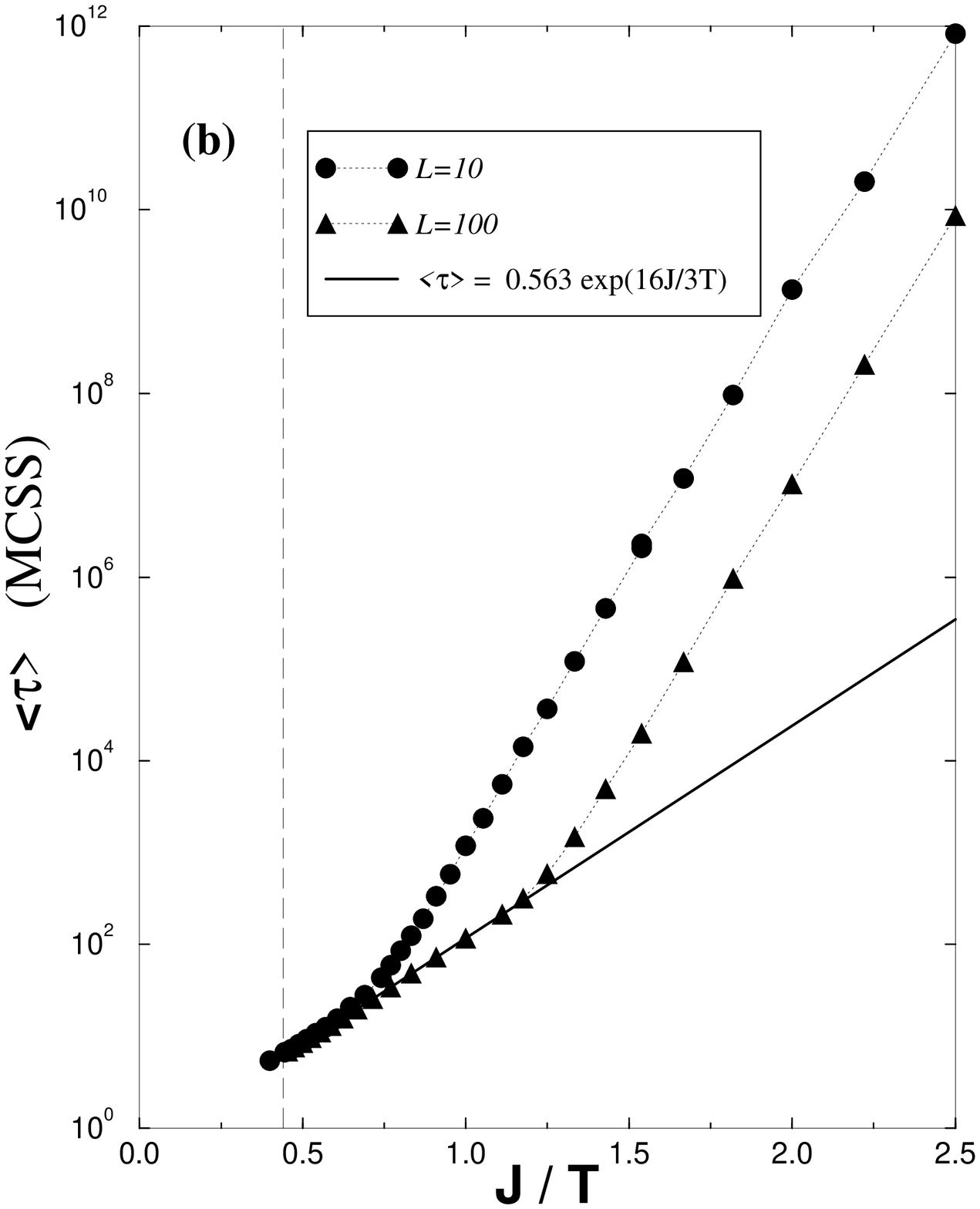}
}
\vspace*{10pt}
\fcaption{
The average lifetimes, $\langle\tau\rangle$, in units of MCSS, 
for $L$$=$$10$ and $L$$=$$100$ are shown 
as functions of inverse temperature for $H$$=$$-0.75$$J$.  
The same data are plotted in (a) and (b).  
The vertical dashed line is $J$$/$$T_{\rm c}$.  
In (a) $L^2$$\langle\tau\rangle$ is plotted and the 
data points lie on top of each other in the SD regime.
In (b) the same data points lie on top of each other 
when $\langle\tau\rangle$ is in the MD and SF regimes.  
The solid lines are from $T$$\rightarrow$$0$ predictions, as 
described in the text.}
\vspace*{13pt}
\end{figure}
As seen in Fig.~2, $\langle\tau\rangle$, 
with units of Monte Carlo steps per spin (MCSS), 
can vary over more than ten orders of magnitude.  
The remainder of the paper describes briefly 
the algorithms that allow such simulations to be performed.  
The data points in Fig.~2 were taken with different techniques 
described in this paper, including the 
Monte Carlo with Absorbing Markov Chains (MCAMC) algorithm (with up to 
$s$$=$$3$) and the Projective Dynamics (PD) algorithm with and without 
a moving wall.  With proper use of the algorithms, 
one obtains the correct lifetimes.  Consequently, which points were 
taken with which algorithms is not indicated in the figure.  
First note that the same data are plotted in parts (a) and (b) of Fig.~2.  
In the SD regime, Fig.~2(a), the lifetimes for 
different $L$ lie on top of each other when $L^2$$\langle\tau\rangle$ 
is plotted.  
In the MD regime, Fig.~2(b), the values of $\langle\tau\rangle$ 
for different $L$ 
lie on top of each other.  
The solid lines in Fig.~2 are 
$T$$\rightarrow$$0$ predictions for the exponential 
dependence valid for $2/3$$\le$$|H|$$/$$J$$\le$$1$ 
from low-temperature predictions\cite{LowT1,LowT2}.  
In the SD regime\cite{LowT1} 
$\langle\tau\rangle$$=$$A_{\rm SD}$$\exp[(24J$$-$$14|H|)/T]$$/$$L^2$ 
and in the 
MD regime\cite{LowT2} to 
$\langle\tau\rangle$$=$$A_{\rm MD}$$\exp[(28J$$-$$16|H|)/3T]$.  
Only the prefactors are fit to the data.  
The SD prefactor $A_{\rm SD}$$=$$0.186$ 
is set to agree with the data 
at $T$$=$$0.4$$J$, 
while the MD prefactor $A_{\rm MD}$$=$$0.563$ 
is set to agree with the 
$L$$=$$100$ data point at $T$$=$$J$.  

\medskip
\setcounter{section}{2}
\setcounter{equation}{0}
\section{Constrained Transfer Matrix}
\noindent 
In the 1960's Langer\cite{Lang60s} used field-theoretical arguments 
to show that the nucleation rate, and hence the lifetime, can be 
related to the imaginary part of the analytically continued free energy 
density, ${\cal I}m({\cal F})$.  In particular, 
$\langle\tau\rangle$$\propto$$\left[{\cal I}m({\cal F})\right]^{-1}$.
Numerically it is possible to obtain estimates for 
${\cal I}m({\cal F})$ by finding all eigenvectors and 
eigenvalues of the $2^\ell$$\times$$2^\ell$ 
transfer matrix for an $\ell$$\times$$\infty$ lattice.  
This is accomplished by introducing constrained entropy 
densities\cite{Chris93}
in a manner analogous to the source entropy of a stationary 
ergodic Markov information source.  
The constrained transfer matrix (CTM) calculations 
obtain imaginary parts of the free energy density from the 
imaginary parts of the constrained entropy density.  
Although the transfer matrix does not contain any explicit 
dynamics, the CTM method 
gives ${\cal I}m({\cal F})$ values that agree\cite{Chris93} with 
droplet predictions for the functional form of $\langle\tau\rangle$.  
It also gives 
values for metastable magnetizations which agree 
with those obtained from Avrami analysis of MD growth\cite{Ramos99}.  
Fig.~3 shows the stable and metastable magnetizations from 
CTM calculations at $|H|$$=$$J$$/$$2$ for various strip widths $\ell$.  
It illustrates some of the strengths and weaknesses of the CTM method.  
\begin{figure}
\vspace*{13pt}
\vspace*{0.4in}
\centerline{
\epsfxsize=3.8in
\epsfysize=3.8in
\epsfbox{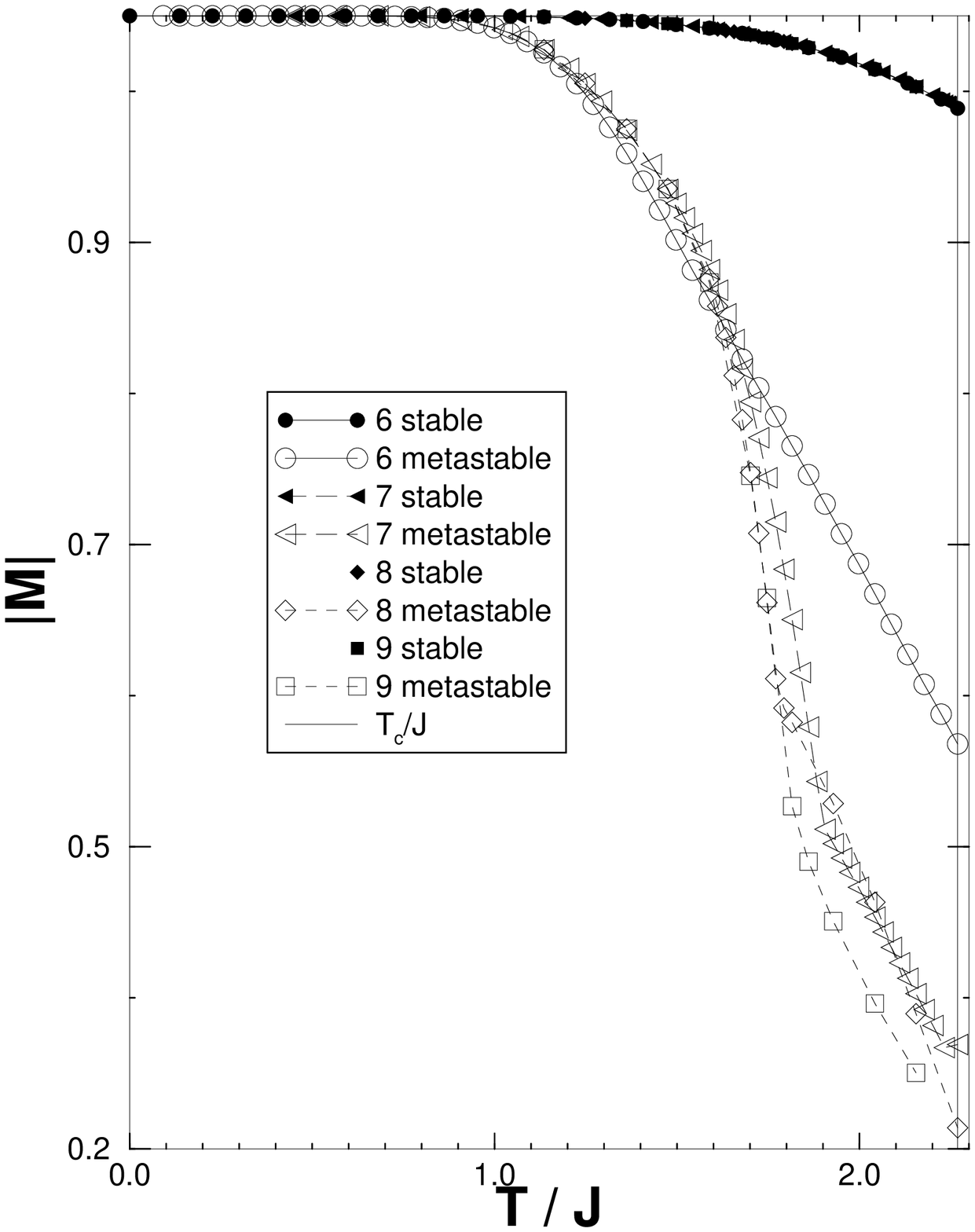}
}
\vspace*{10pt}
\fcaption{The absolute values of the 
stable (closed symbols) and metastable magnetizations 
(open symbols) 
at $H$$=$$-0.5$$J$ for $\ell$$=$$6, 7, 8, 9$.  
Note that the crossover between the SD and MD regimes is seen 
near $T$$=$$1.75$$J$.}
\vspace*{13pt}
\end{figure}
\begin{itemlist}
\vskip 0.0 true cm
\item CTM strengths
\vskip 0.0 true cm
\begin{itemlist}
\vskip 0.0 true cm
\item[$\circ$] Can obtain ${\cal I}$$m$$($${\cal F}$$)$ directly.  
\vskip 0.0 true cm
\item[$\circ$] Can obtain metastable magnetization easily.  
\vskip 0.0 true cm
\end{itemlist}
\vskip 0.0 true cm
\item CTM weaknesses
\vskip 0.0 true cm
\begin{itemlist}
\vskip 0.0 true cm
\item[$\circ$] Only feasible for small strip widths.  This restricts 
applications to either mean-field like models or $d$$=$$2$ models 
with short-range interactions.  
\vskip 0.0 true cm
\item[$\circ$] Must use very high precision in the computer calculations.  
\vskip 0.0 true cm
\item[$\circ$] Interpretation subtle because of `lobes' in 
transfer-matrix spectrum.  
\vskip 0.0 true cm
\end{itemlist}
\end{itemlist}

\medskip
\setcounter{section}{3}
\setcounter{equation}{0}
\section{Monte Carlo with Absorbing Markov Chains}
\noindent 
The first algorithm to obtain very large increases in simulated times for 
faithful dynamics of 
discrete spin models was put forward about 25~years ago by Bortz, 
Kalos, and Lebowitz\cite{nfold75}.  This $n$-fold way method 
is extremely efficient for low-$T$ simulations.  
The $n$-fold way algorithm has been rewritten\cite{mynfold}
for the discrete-time algorithm used in this paper, and 
this version is presented here.  
The underlying dynamics corresponds to a 
$2^{L^2}$$\times$$2^{L^2}$ Markov matrix.  Since this matrix is too large 
to diagonalize\cite{nightnow}, we can deal with it one piece at a 
time within our Monte Carlo simulation.  If we look at an 
$s$$\times$$s$ submatrix ${\bf T}$ of the original Markov matrix, 
one has a matrix associated with the absorbing Markov chain (AMC) 
\begin{equation}        
{\bf M} = 
\left (
\begin{array}{ll}
{\bf I}_{q\!\times\!q} & {\bf 0}_{q\!\times\!s} \\
{\bf R}_{s\!\times\!q} & {\bf T}_{s\!\times\!s}
\end{array}
\right )
\end{equation}
where ${\bf R}$ takes into account the $q$ ways to get 
out of the $s$ states included in the transient matrix ${\bf T}$.  
Here ${\bf I}$ is the identity matrix, and ${\bf 0}$ is the matrix with 
all zero elements.  
The initial vector ${\vec v}_{\rm I}^{\rm T}$ 
(in this section we use the convention common in mathematics 
where operators act to the left) 
has zeroes everywhere with the 
exception of a one in the current spin configuration, which must be one 
of the $s$ transient configurations.  
The random time $m$ for exiting from the $s$ transient states 
is the solution of the equation 
\begin{equation}	
{\vec v}_{\rm I}^{\rm T} {\bf T}_{s\!\times\!s}^m{\vec e}
< r \le 
{\vec v}_{\rm I}^{\rm T} {\bf T}_{s\!\times\!s}^{m\!-\!1}{\vec e}
\end{equation}
where 
$r$ is a uniformly distributed random number between $0$ and $1$, 
and ${\vec e}$ is the vector with all elements unity.  
Given that the $s$ transient states are exited at time-step $m$, 
the vector of (unnormalized) probabilities to exit to a particular 
one of the $q$ states is given by 
\begin{equation}	
{\vec{\cal Q}}^{\rm T} = 
{\vec v}_{\rm I}^{\rm T} {\bf T}_{s\!\times\!s}^{m\!-\!1}
{\bf R}_{s\!\times\!q}
.
\end{equation}
Once the $s$ transient states have been exited, a new 
transient subspace is chosen and the algorithm is repeated.  
It is only Eqns.~(3.2) and (3.3) which need to be used to incorporate 
the methodology of AMC into a Monte Carlo 
simulation\cite{MCAMCuga,MCAMCprl}.  

For the isotropic square-lattice Ising model 
in a field, each spin belongs to 
one of $n$$=$$10$ classes.  
The first $5$ classes ($1$$\le$$i$$\le$$5$) have spin up and 
$5$$-$$i$ nearest-neighbor spins which are up.  
The last $5$ classes ($6$$\le$$i$$\le$$10$) have spin down and 
$10$$-$$i$ nearest-neighbor spins which are up.  
Let $c_i$ be the number of spins in class $i$, and let $p_i$ be the 
probability of flipping a spin in class $i$ given that that particular 
spin was chosen for a spin-flip attempt.  

If $s$$=$$1$, only the current spin configuration is in the 
transient subspace.  This corresponds to the discrete-time 
$n$-fold way algorithm.  In that case, for our square-lattice 
Ising model, the transient matrix is 
${\bf T}_{1\!\times\!1}$$=$$1-L^{-2}\sum_{i=1}^{10} c_i p_i$$=$$1 - 
L^{-2} Q_{10}$, 
which defines $Q_{10}$.  Then Eq.~(3.2) becomes 
$m$$-$$1$$\le$$\ln(r)$$/$$\ln($$1$$-$$L^{-2}$$Q_{10}$$)$$<$$m$.
For $Q_{10}$$\ll$$L^2$ the time of exiting can be 
approximated by\cite{mynfold} 
$m$$\approx$$\Delta t$$=$$-L^2$$\ln(r)$$/$$Q_{10}$
which leads to the continuous-time version of the 
$n$-fold way algorithm.  
The probability of flipping 
one of the spins in class $i$ is given from Eq.~(3.3) by 
$c_i p_i / Q_{10}$, now independent of $m$.  
\begin{itemlist}
\vskip 0.0 true cm
\item MCAMC strengths
\vskip 0.0 true cm
\begin{itemlist}
\vskip 0.0 true cm
\item[$\circ$] Can give {\bf MANY\/} orders of magnitude speed-up, so 
extremely large lifetimes can be simulated.  
\vskip 0.0 true cm
\item[$\circ$] Extremely efficient at low temperatures and small $L$, 
{\it i.e.,} in the SD regime.  
\vskip 0.0 true cm
\item[$\circ$] Can have smaller errors in values for $\langle\tau\rangle$ 
compared with the same number of metastable escapes using normal simulations.  
This is because the exact average lifetime for exiting from the $s$ transient 
states is given by 
${\vec v}_{\rm I}^{\rm T} ({\bf I}-{\bf T})^{-1}{\vec e}$, which can be 
incorporated into $\langle\tau\rangle$.  
\vskip 0.0 true cm
\item[$\circ$] Can calculate higher moments of the escape time.  
\vskip 0.0 true cm
\item[$\circ$] Provides a direct connection between discrete-time 
Monte Carlo simulations and the continuous time used in the 
original $n$-fold way algorithm.  
\vskip 0.0 true cm
\item[$\circ$] Can be parallelized in a 
non-trivial fashion, which has been done for $s$$=$$1$\cite{George99}.  
This allows one to efficiently use 
multi-processor scalable machines to simulate 
discrete-event processes on large lattices for long times.  
\vskip 0.0 true cm
\end{itemlist}
\vskip 0.0 true cm
\item MCAMC weaknesses
\vskip 0.0 true cm
\begin{itemlist}
\vskip 0.0 true cm
\item[$\circ$] Requires tabulation of Markov sub-matrices, which can 
be complicated for large $L$ or small $H$.  
\vskip 0.0 true cm
\item[$\circ$] Requires a small number of spins that evolve relatively 
quickly, so system stays in Markov sub-matrix for a large number of 
spin flip attempts.  
\vskip 0.0 true cm
\item[$\circ$] Easiest to program efficiently for a small number of 
spin classes.  
\vskip 0.0 true cm
\item[$\circ$] The parallelized $s$$=$$1$ version looses some of 
its efficiency at low temperatures, compared with 
the serial algorithm\cite{George99}.  
\vskip 0.0 true cm
\end{itemlist}
\vskip 0.0 true cm
\end{itemlist}

\medskip
\setcounter{section}{4}
\setcounter{equation}{0}
\section{Projective Dynamics}
\noindent
In order to show the similarities between the Projective Dynamics 
(PD) method\cite{PDuga,PDprl} and the 
Transition Matrix Monte Carlo (TMMC) method\cite{Wang99}
we use the TMMC notation in this section.  The 
dynamics of the single-spin flip process acting on the 
$2^{L^2}$ state vector $P$
is given by the 
continuous-time Markov matrix $\Gamma$, 
\begin{equation}	
{{\partial P(\sigma,t)}\over{\partial t}} =
\sum_{ \{ \sigma'\} } \Gamma(\sigma|\sigma') P(\sigma',t) 
.  
\end{equation}
The idea behind both the PD and TMMC methods is to coarse-grain this 
process by lumping together some of the $2^{L^2}$ states according 
to some variable $C$.  The probability of going from lumped state $C$ 
to $C'$ is given by 
\begin{equation}	
W(C|C') = {1\over{n(C')}} 
\>\sum_{ \sigma\in\{\sigma|C\} }
\>\sum_{ \sigma'\in\{\sigma|C'\} }
\>\Gamma(\sigma|\sigma')
\end{equation}
where $n(C')$ is the number of original states lumped 
into state $C'$.  
This yields 
\begin{equation}	
{{\partial P(C,t)}\over{\partial t}} =
\sum_{ \{ C'\} } W(C|C') P(C',t) 
\end{equation}
where $P(C,t)$ is the probability of being in lumped state $C$ at time $t$. 
For the PD method we choose $C$$=$$M$, so we lump on the magnetization, 
and since we are interested in metastable decay we work at finite field 
and temperatures well below $T_{\rm c}$.  
In the TMMC\cite{Wang99} one lumps on the energy, so $C$$=$$E$, and one is 
interested in the critical behavior and works at $H$$=$$0$ and 
temperatures near $T_{\rm c}$.  
One interesting result is that the TMMC method yields 
dynamics much different\cite{Wang99} 
from single spin-flip dynamics, due to the 
allowed `microcanonical' mixing within the lumped states.  
However, lumping on $M$ does not allow spin flips to occur without 
jumping out of the current lumped state --- and it has been 
proven that within statistics the correct value of $\langle\tau\rangle$ is 
obtained\cite{PDexact}.  This means that memory effects due to the 
lumping in $M$ do not enter into the calculation of $\langle\tau\rangle$.  
For the PD case, the lumped matrix is tridiagonal and one has 
$W(M|M')$$=$$s(M') \delta_{M+2,M'} $$+$$g(M') \delta_{M-2,M'}$ 
which defines the growing and shrinking rates of the stable phase as 
$g(M)$$=$$\sum_{i=1}^5 \langle c_i\rangle_M p_i$ and 
$s(M)$$=$$\sum_{i=6}^{10} \langle c_i\rangle_M p_i$.  
Here $\langle c_i\rangle_M$ is the average number of spins in class $i$ 
during the simulated escape 
given that the configuration has magnetization $M$.  

The random walk starts at $M$$=$$L^2$ and terminates 
at $M$$=$$0$.  
We define $h(M)$ as the total time spent in $M$.  
If $L$ is even, 
$h(M)$$=$$0$ for $M$ odd, since these magnetizations are 
not possible.  Then 
\begin{equation}	
\langle\tau\rangle = \sum_{M=2}^{L^2} h(M), \>\>\>
h(M) = {{1+s(M-2)h(M-2)}\over{g(M)}}, \>\>\> 
h(2)= {{1}\over{g(2)}}
.  
\end{equation}

It is also possible to introduce a moving wall in the magnetization 
to force the system out of the metastable state, and to measure 
$g(n)$ and $s(n)$ during this (hopefully quasi-static) process.  
The wall position is given by 
$M_{\rm wall}(t)$$=$$(L^2\!+\!1)$$-$$v_{\rm wall}  t$.  
For a soft wall the normal Monte Carlo flip probability is 
multiplied by 
$p_{\rm wall}$$=$$\exp\left[-c\left(M_{\rm new}-M_{\rm wall}\right)\right]$
if the wall at position $M_{\rm wall}$ is past the 
magnetization of the Monte Carlo new trial configuration.  
Another possibility is to introduce a hard wall, $c$$=$$\infty$, that 
always flips a chosen up spin if $M_{\rm wall}$ is past the current 
magnetization.  Fig.~4 shows an example of 
lifetimes for different forcing walls 
as functions of the wall speed.  The difference between the lifetimes 
measured from Eq.~(4.4) 
and the number of MCSS before escape with the wall 
gives an estimate for the speed-up due to incorporating the 
forcing wall.  Of course, for a wall moving too fast the process is 
not quasi-static and the lifetimes are not accurate.  However, 
good results are obtained even for relatively fast walls.  For example 
for the hard wall a velocity of $3$$\times$$10^{-4}$M/$L^2$MCSS gives 
a lifetime comparable to an actual escape but requires almost 
two orders of magnitude less computer time.  
The soft walls do not seem to help the convergence, and 
additional calculations are needed to calculate $p_{\rm wall}$.  
This makes the hard wall more 
computationally efficient compared with the soft walls.  
\begin{figure}
\vspace*{0.7in}
\centerline{
\epsfxsize=3.8in
\epsfysize=3.8in
\epsfbox{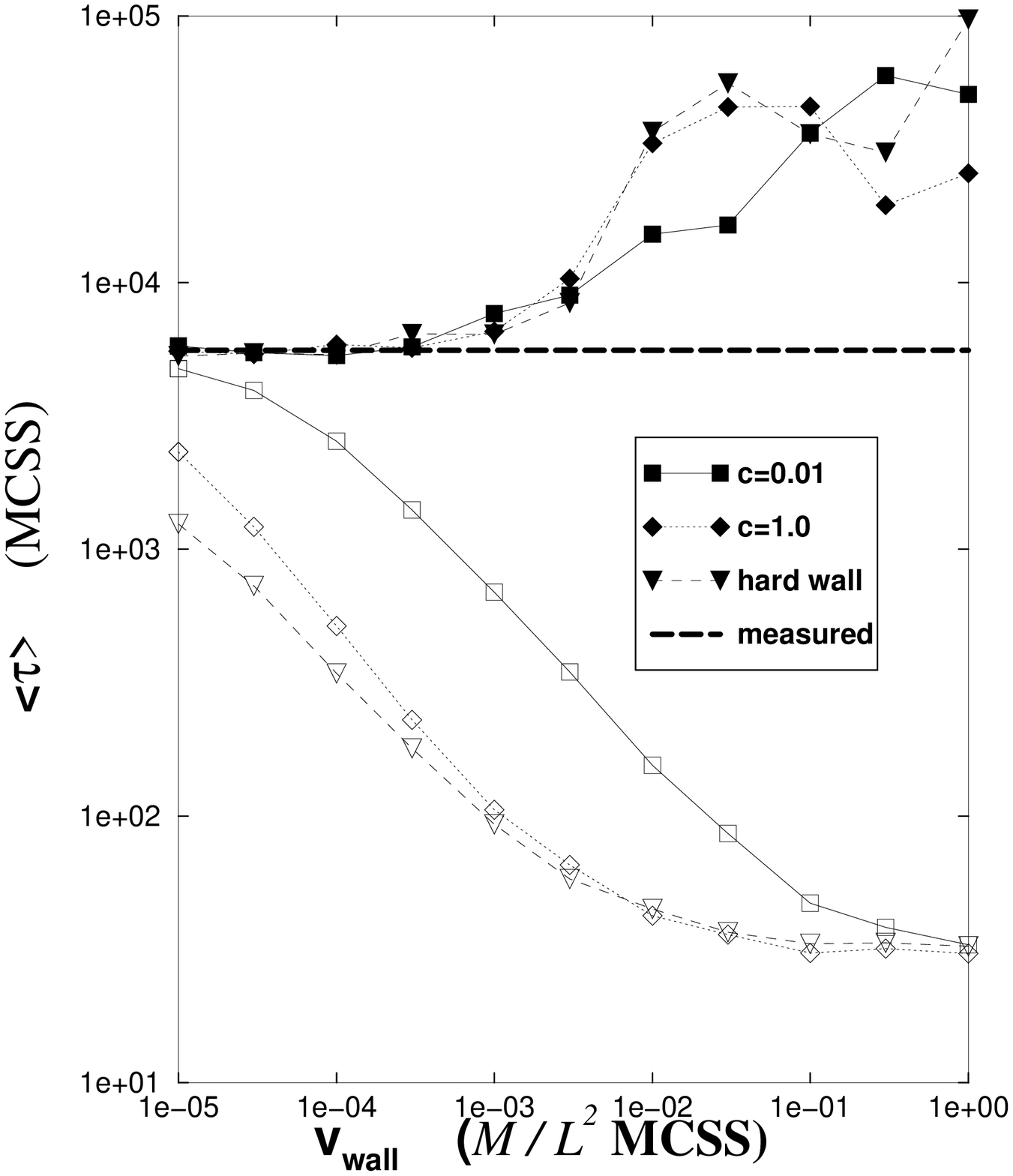}
}
\vspace*{10pt}
\fcaption{The average lifetime, $\langle\tau\rangle$, in units 
of MCSS is shown as a function of the velocity of the 
forcing wall (in units of $M$$/$$L^2$MCSS).  This is for a 
$10$$\times$$10$ lattice with $T$$=$$0.9$$J$ and $H$$=$$-0.75$$J$.  
The data are for different degrees of softness of the wall, 
$c$$=$$0.01$ ($\Box$), 
$c$$=$$1.0$ ($\Diamond$), 
and a hard wall $c$$=$$\infty$ ($\bigtriangledown$).  
The filled symbols are $\langle\tau\rangle$ values from the 
growing and shrinking probabilities, while the corresponding 
open symbols are the measured number of MCSS to escape 
from the metastable state with the forcing wall.  
The dotted lines connect the data points and serve as guides to the eye.  
The dashed horizontal line corresponds to the result for $\langle\tau\rangle$ 
from standard simulations.  
}
\end{figure}
\begin{itemlist}
\vskip 0.0 true cm
\item PD strengths
\vskip 0.0 true cm
\begin{itemlist}
\vskip 0.0 true cm
\item[$\circ$] Can give {\bf MANY\/} orders of magnitude speed-up, so 
extremely large lifetimes are possible.  
\vskip 0.0 true cm
\item[$\circ$] Easy to obtain magnetization of metastable 
state\cite{PDuga,PDmrs}.  
\vskip 0.0 true cm
\item[$\circ$] Easy to obtain magnetization of saddle 
point\cite{PDuga,PDmrs}.  
\vskip 0.0 true cm
\item[$\circ$] Allows extrapolation to very long lifetimes for weak 
$H$\cite{PDuga}.  
\vskip 0.0 true cm
\item[$\circ$] Allows extrapolation to large $L$ from simulations at 
smaller $L$\cite{PDuga}.  
\vskip 0.0 true cm
\item[$\circ$] Gives {\bf EXACT\/} value of lifetime, except for 
statistical deviations\cite{PDexact}.  
\vskip 0.0 true cm
\item[$\circ$] Has decreased errors in values for $\langle\tau\rangle$ 
compared with same number of metastable escapes using normal simulations.  
\vskip 0.0 true cm
\item[$\circ$] Works well in both SD and MD regimes, as well as 
the coexistence regime.  
\vskip 0.0 true cm
\item[$\circ$] Can incorporate a `moving forcing wall' to further 
enhance performance\cite{PDprl}.  
\vskip 0.0 true cm
\end{itemlist}
\vskip 0.0 true cm
\item PD weaknesses
\vskip 0.0 true cm
\begin{itemlist}
\vskip 0.0 true cm
\item[$\circ$] Gives only approximate higher moments of lifetime 
due to memory effects ignored when the states are lumped\cite{PDuga}.  
\vskip 0.0 true cm
\item[$\circ$] The original Markov chain is not `lumpable', nor 
is it even `weakly lumpable'\cite{PDprl}.  
\vskip 0.0 true cm
\item[$\circ$] It is nontrivial to know how fast to move the wall.  
\vskip 0.0 true cm
\end{itemlist}
\end{itemlist}

\medskip
\section{Discussion and Conclusions}
\noindent
We have demonstrated several dynamically faithful 
algorithms to study homogeneous nucleation and 
growth in discrete spin models.  These include 
the Constrained Transfer Matrix (CTM) method, 
the Monte Carlo with Absorbing Markov Chains (MCAMC) 
method which is a generalization of the $n$-fold way algorithm, 
and the 
Projective Dynamics (PD) method.  Each method has its 
own strengths and weaknesses, which are detailed 
in Sec.~2, 3, and 4, respectively.  
Consequently, which method should be used 
depends on the model to be studied and on the parameter 
regime where the model is to be studied.  
Application and generalization of these methods to 
faithfully study the dynamics of other systems, 
including systems with randomness, 
continuous spin models\cite{UsHeis}, 
higher-dimensional systems\cite{PDuga,PDprl}, 
and 
systems evolving via stochastic differential equations 
are some possible future interesting applications.  

\nonumsection{Acknowledgments}
\noindent
The author would like to thank 
M.~Kolesik, 
M.P.\ Nightingale, 
P.A.\ Rikvold, 
and 
R.H.\ Swendsen, 
for useful discussions.  
Supported in part by the NSF through grant number DMR-9871455,
and through the Supercomputer Computations Research Institute which is
funded by the U.S.\ DOE and the State of Florida.  
Supercomputer time supplied by 
National Energy Research Scientific Computing Center (NERSC).  

\nonumsection{References}

\end{document}